\documentclass[a4paper,10pt]{article} 
\usepackage{cite}
\usepackage{setspace}
\usepackage{ae,aecompl,amsbsy,amssymb,eurosym}
\usepackage[english]{babel}
\usepackage{graphicx}
\usepackage{epsfig}
\usepackage{epstopdf}
\usepackage{amsmath}
\usepackage{amssymb}
\usepackage{subfigure}
\usepackage{ifthen}
\usepackage{alltt}
\usepackage{fancyhdr}
\usepackage{verbatim}
\usepackage{moreverb}
\usepackage{url}

\newcommand{\vt}[1]{\ensuremath{\mathbf{#1}}} 
\newcommand{\lt}[1]{\ensuremath{\mathrm{#1}}} 

\fancypagestyle{plain}{
  \fancyhead{} 
}

\begin{document}

\author{Antti~O.~Karilainen$^1$, Pekka~M.~T.~Ikonen$^2$, Constantin~R.~Simovski$^1$,\\ Sergei~A.~Tretyakov$^1$, Andrey~N.~Lagarkov$^3$, Sergei~A.~Maklakov$^3$,\\ Konstantin~N.~Rozanov$^3$, and Sergey~N.~Starostenko$^3$\\
$^1$Department of Radio Science and Engineering/SMARAD Center of Excellence,\\ Helsinki University of Technology (TKK),\\ P.O.Box 3000, FI-02015 TKK, Finland.\\ Email: (see http://radio.tkk.fi/en/contact/).\\
$^2$TDK-EPC, P.O.Box 275, FI-02601 Espoo, Finland. Email: pekka.ikonen@epcos.com.\\
$^3$Institute for Theoretical and Applied Electromagnetics (ITAE),\\ Russian Academy of Sciences,\\ 13 Izhorskya ul., 125412 Moscow, Russia.\\ Email: Konstantin Rozanov krozanov@yandex.ru.}

\title{Experimental Studies of Antenna Miniaturization Using Magneto-Dielectric and Dielectric Materials}

\maketitle

\begin{abstract}
Measurement results for a meandered planar inverted-F antenna (PIFA) loaded with magneto-dielectric and dielectric materials are presented. Figures of merit and ways to compare antennas with different fillings materials are discussed. The used magneto-dielectric material is described, the radiation mechanism of the meandered PIFA is studied, and the proper position for dielectric and magneto-dielectric filling is discussed and identified. Identical-size antennas with dielectric and magneto-dielectric fillings are compared at the same resonance frequency using the radiation quality factor as the figure of merit. It is seen, that the benefit from the magneto-dielectric filling material is moderate and strongly dependent on the positioning of the filling.
\end{abstract}


\section{Introduction}

A substrate with lossless and dispersion-free permeability $\mu$ higher than the permittivity $\epsilon$ could theoretically lead to substantially wider impedance bandwidths in patch antennas \cite{Hansen2000}. However, new artificial and composite magnetic materials have not yet proved their expectations in patch-antenna miniaturization. Artificial magnetic materials, such as those based on split-ring resonators, are dispersive by nature, and in case of Lorentzian-type dispersion, patch-antenna miniaturization in terms of improved radiation quality factor is impossible, as compared to a reference design \cite{Ikonen2006b}. As opposed to artificial materials, substrates with natural magnetic inclusions can provide reasonably large static values of permeability with relatively low dispersion, and miniaturization is possible \cite{Ikonen2006c}. In this paper we present measurement results of a meandered planar inverted-F antenna (PIFA) miniaturized with a magneto-dielectric material, and miniaturization using conventional dielectric material is used as a reference for comparison.

Modern material manufacturing technology has made possible to design composite substrates with magnetic inclusions mixed with dielectric host materials. Antennas with magneto-dielectric materials are studied experimentally in \cite{Tanaka2004,Min2006,Deng2007,Sun2007,He2008,Yang2008,Yang2008b,Petrov2008,Petrov2008b,Shin2008}. Known magnetic materials tend to be lossy, and in antenna applications the radiation efficiency becomes the main figure of merit in addition to the unloaded quality factor or impedance bandwidth. Previous results in antenna miniaturization and efficiency include \cite{Bae2008,Grange2009,Yang2009}.

When using magneto-dielectric substrates in miniaturization, miniaturization is equivalent to a decrease of  the resonant electrical size of the antenna at a given operating frequency. A question that should be answered when using novel materials is whether the used material outperforms the traditional miniaturization methods \cite{Ikonen2009}. If we accept that by using magnetic substrates we can, at least in theory, miniaturize a patch antenna without decreasing noticeably the bandwidth, we also have to challenge it against available dielectric material. Good-quality dielectric materials have low loss, but magneto-dielectric materials are typically considerably more lossy. Therefore, both the radiation efficiency and the measured unloaded quality factor must be used when comparing the same antenna with magneto-dielectric and dielectric fillings. This has been done in this paper by using the radiation quality factor as as a figure of merit, where the effect of dissipative losses in the antenna have been normalized away from the results.

The quality factor of a small antenna depends on the operating frequency and the antenna size in addition to effective utilization of the antenna's volume \cite{balanis}, so one has to be careful when comparing antennas with different fillings. When comparing an antenna with magneto-dielectric and dielectric materials, the quality factor must be measured at the same frequency for both cases. Moreover, the physical size of the antenna should also be the same. We discuss in this article the necessary conditions for fair comparison between antennas with different material fillings.

Resonant antennas in general behave differently when using magnetic or dielectric materials for filling. This can be understood easily from the field distributions inside antennas. It has been proposed, that in some cases the optimal filling material can be determined from the radiating fields \cite{Karilainen2009,Karilainen2009e}. We discuss how to choose and also position the used filling materials for antennas in general, and use these guidelines for the meandered PIFA under study~\cite{Karilainen2009d}.

In Section~\ref{sec:antennas}, we describe the needed figures of merit and discuss how to compare antennas in practice. Section~\ref{sec:material} describes the used magneto-dielectric material and Section~\ref{sec:experiment} presents the antenna under test and the results of the measurements. The measurement results are analyzed and discussed in Section~\ref{sec:discussion} and conclusions are made in Section~\ref{sec:conclusion}.


\section{Antenna Merits and Comparison}
\label{sec:antennas}

Before we start describing the measurements, we will review and discuss the measured parameters. The obvious figure of merit for small antenna measurements is the unloaded quality factor, or the radiation quality factor when the radiation efficiency of the antenna is included in calculations. However, when comparing two small antennas, we must also consider the size or volume, and the resonance frequency together.


\subsection{Figures of Merit for Small Antennas}

First we discuss the figures of merit for antenna measurements. In other words, we define \emph{what} we want to use when comparing performance of antennas. The radiation efficiency is the ratio between the radiated power $P_\lt{r}$ and the total accepted power $P_\lt{tot}$. $P_\lt{tot}$ is the sum of $P_\lt{r}$ and the dissipated power $P_\lt{L}$ in the antenna. Using the Thevenin equivalent antenna impedance $Z_\lt{a} = R_\lt{a} + jX_\lt{a}$ and dividing the antenna resistance as $R_\lt{a} = R_\lt{r} + R_\lt{L}$, where $R_\lt{r}$ corresponds to radiated power and $R_\lt{L}$ to dissipated power, we can write the radiation efficiency as (e.g.~\cite{balanis}):
\begin{equation}
  \eta_\lt{r} = \frac{P_\lt{r}}{P_\lt{tot}} = \frac{P_\lt{r}}{P_\lt{r} + P_\lt{L}} = \frac{R_\lt{r}}{R_\lt{r} + R_\lt{L}}.
  \label{eq:radeff}
\end{equation}
Using the Wheeler-cap method to calculate the radiation efficiency \cite{Wheeler1959}, the right-hand part of (\ref{eq:radeff}) can be directly applied to find $\eta_\lt{r}$ from the measured $Z_\lt{a}$.

The antenna quality factor $Q$ and the bandwidth $B$ of an antenna with a single resonance are related to each other with the voltage standing wave ratios (VSWRs) as follows: If the bandwidth criterion is $\lt{VSWR}<S$ and $T$ is the VSWR at the resonant frequency, $Q$ can be calculated as~\cite{Pues1989}
\begin{equation}
  Q = \frac{1}{B}\sqrt{\frac{ (TS - 1)(S-T) }{ S }}.
  \label{eq:QB}
\end{equation}
It follows from (\ref{eq:QB}), that for a certain desired matching level we should have half the reflection coefficient for the optimal \emph{over-coupled} resonance. For example, for $S_{11}=-6$~dB matching level, the reflection coefficient should be $-12$~dB at the center frequency of the resonance.

It has been recently proposed, that the quality factor can be approximatively solved directly from the impedance also for antennas with \emph{lossy and dispersive} materials:~\cite{Yaghjian2005,Yaghjian2006}
\begin{equation}
	Q = \frac{\omega}{2R_\lt{a}} \bigg| \frac{\partial Z_\lt{a}}{\partial \omega} \bigg|_{\omega = \omega_0},
	\label{eq:QZ}
\end{equation}
although, strictly speaking, the knowledge of only $Z_\lt{a}$ is not enough to find the ratio between radiated, stored and dissipated powers.

When the quality factor is calculated from (\ref{eq:QB}) or (\ref{eq:QZ}), the effect of dissipated power in the antenna can be removed from $Q$ by using the radiation quality factor
\begin{equation}
	Q_\lt{r} = \frac{Q}{\eta_\lt{r}}.
	\label{eq:Qr}
\end{equation}
In practice, when we measure an antenna and solve $Q$ from the $S_{11}$ parameter or from $Z_\lt{a}$, we must use (\ref{eq:Qr}) to compare antennas with different $\eta_\lt{r}$.


\subsection{Antenna Filling}

Now that we know what to compare in antenna measurements, we must set up rules \emph{how} we compare antennas with different loading materials. When dealing with small antennas, the well-known lower limit for the quality factor (e.g.~\cite{balanis})
\begin{equation}
	Q \geq \frac{1}{(ka)^3},
	\label{eq:chu_limit}
\end{equation}
can be thought to hold with $ka \ll 1$. It is calculated from the stored average energy outside a sphere with radius $a$ surrounding the antenna. If we assume that $a^3$ is proportional to the volume $V$ of an antenna, we can write the minimum quality factor as follows:
\begin{equation}
	Q \sim \frac{1}{f^3V},
	\label{eq:limit_fv}
\end{equation}
since $k=2\pi f\sqrt{\mu\epsilon}$. Now, it is obvious from (\ref{eq:limit_fv}) that if we compare the quality factors of two antennas, they should have the same volume $V$ \emph{and} the same operating frequency $f$. Otherwise it is not possible to come to any conclusion which one of the antennas has e.g.\ the largest potential bandwidth and comes close to the fundamental limit. In fact, the bandwidth itself is a difficult figure of merit since both $Q$ and the radiation efficiency affect it. Because of this, we favor the radiation quality factor below.

The effect of material filling, partial or full, on antenna performance is not simple. Is has been proposed in \cite{Karilainen2009} that the most beneficial material type can be determined analyzing the radiating currents of the antenna. Next, we summarize the steps needed for an efficient design of a small material-loaded antenna.

\begin{itemize}
	\item[i)] Identify the radiating fields or currents. For example, in case of patch antennas, most of the radiation is produced by the fringing electric fields at the open ends of the patch. The surface current due to the magnetic field is canceled by the image current below the ground plane, but the magnetic equivalent current from the electric field is doubled.
	
	\item[ii)] Determine the fields contributing mainly to the stored energy. If either electric or magnetic field does not provide radiation, it is safe to use dielectric or magnetic materials, respectively, to suppress these fields and make the resonating wavelength larger, i.e.\ to miniaturize the antenna.
		
	\item[iii)] Apply the filling to the right place. If possible, use the filling material only at the place where it has the desired effect. There is no need to use dielectric or magnetic filling in places where there is no electric or magnetic fields, respectively.
\end{itemize}

Magneto-dielectric materials have, of course, both magnetic and dielectric material responses. If we use the material in a position with strong magnetic field but weak electric field, we use only the magnetic properties which may be beneficial and not the harmful dielectric properties, depending on the antenna type. When comparing miniaturization using different material types, the comparison should still be fair. By following the aforementioned rules, magneto-dielectric material should be positioned \emph{only} in the magnetic field, and dielectric material at least in the electric field.

Patch antennas have been studied in \cite{Luukkonen2007,Ikonen2008} with low-loss substrates, and it is seen that the material loading affects not only the bandwidth, but also the current and field distributions inside the antenna cavity and the radiation properties of the antenna. Miniaturization of antennas with with dispersive materials and considerable losses is difficult to analyze, and good analytical models are not available for complicated antennas.


\section{Magneto-Dielectric Composite Material}
\label{sec:material}

The magneto-dielectric material for volumetric antenna filling must be a bulk material if a considerable filling fraction is desired in patch antenna applications. One promising candidate is a magneto-dielectric composite material which is composed of mylar substrate and sputtered multi-layered Fe-SiO$_2$ films \cite{Iakubov2004,Iakubov2007,Iakubov2009}. By using this technology, magnetic response can be achieved at microwave frequencies without excessively high dissipation losses. For the discussion about magnetic materials for microwave applications, see \cite[Section~III]{Ikonen2006c}.

With this technology, bulk magnetic samples of $30 \times 20 \times 0.5$~mm$^3$ in size were manufactured. Each sample consisted of 28 multi-layer films glued each to other. Each film included five Fe-N layers of 0.07~$\mu$m in thickness deposited on both sides of 0.07-$\mu$m thick mylar substrate, with the total of ten layers. The ferromagnetic layers were separated by SiO$_2$ interlayers of $0.07 \mu$m thickness. The thickness of the glue layers between the films was approximately 5~$\mu$m.

\begin{figure}[!t]
  \centering
  \subfigure[]
  {
    \includegraphics[width=50mm]{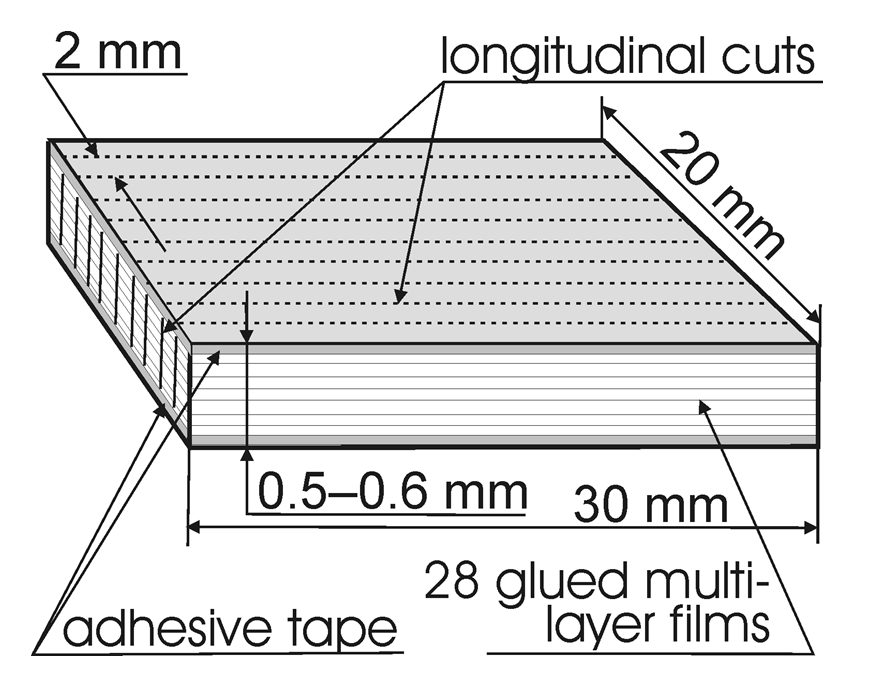}
    \label{fig:material_stack}
  }
  \subfigure[]
  {
    \includegraphics[width=85mm]{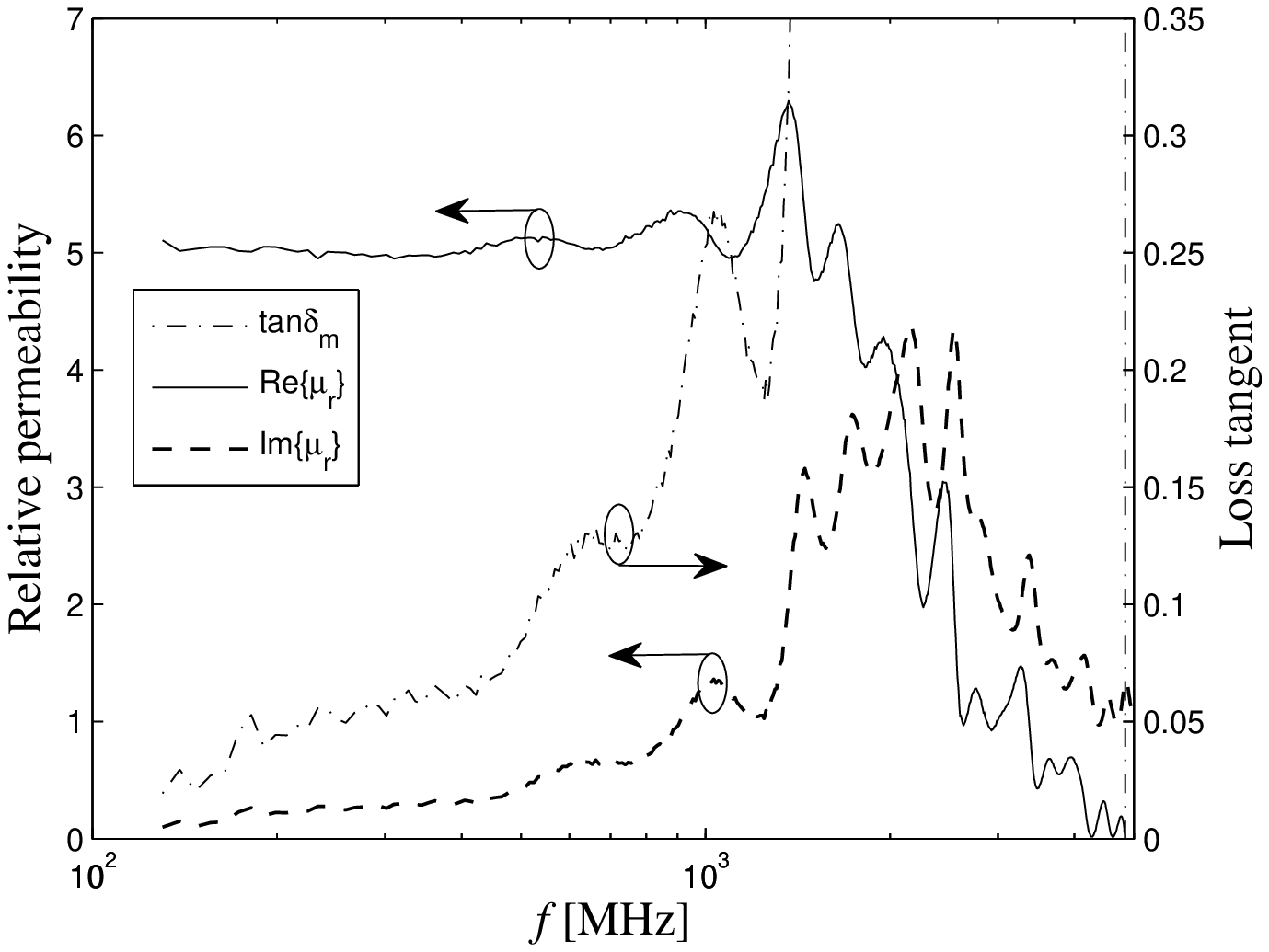}
    \label{fig:mu}
  }
  \caption{a)~Scheme of the sample. b)~Measured typical dispersive relative permeability for the magneto-dielectric material.}
  \label{fig:material}
\end{figure}

To reduce the conductivity along the film plane, the glued samples are cut into stripes of 2 mm in width, stacked together and secured by two pieces of adhesive tape as is shown in the figure. The thickness of the adhesive tape is about 40~$\mu$m, therefore the total thickness of the sampled is between 0.5 and 0.6~mm. The concentration of the magnetic constituent in the sample is 3.5\%. The structure of the sample is sketched in Fig.~\ref{fig:material_stack}.

The microwave performance has been experimentally studied with the strip-line technique \cite{Starostenko2008}. The measured permeability of the sample is presented in Fig.~\ref{fig:mu}, for the frequency range of $130$~MHz to $5$~GHz. The relative permeability e.g.\ at $f=500$~MHz is $\mu_\lt{r} = \mu_\lt{r}'-j\mu_\lt{r}'' = \mu_\lt{r}'(1-j\tan\delta_\lt{m}) = 5.1(1-j0.08)$, where the magnetic loss tangent $\tan\delta_\lt{m}=\mu_\lt{r}''/\mu_\lt{r}'$. The bulk sample was stacked from a series of laminates so that the magnetic response was isotropic along the layers. This in-plane isotropy allows more flexible use of the material in antennas. The relative permittivity of the material is in the neighborhood of $\epsilon_\lt{r}'=2.2 \ldots 2.4$ with the dielectric loss tangent of about $\tan\delta_\lt{e}=0.01$. The in-plane conductivity of the layered material along the cuts is high, which can lead to considerable losses if the electric field exists in this direction.


\section{Experimental Antenna Comparison}
\label{sec:experiment}

One of the antenna types suitable for loading with magnetic materials is the patch antenna. However, when examining the permeability curves of the material seen in Fig.~\ref{fig:mu}, we see that the material becomes very lossy when the frequency is in the gigahertz range. In frequencies below 1~GHz, the wavelength is however so long, that a normal $\lambda/2$ patch antenna is simply too large for e.g.\ portable-device applications. Therefore, we decided to use a planar inverted-F antenna (PIFA) which is a compact version of a shorted $\lambda/4$ patch antenna. The size was reduced even further by meandering the patch of the PIFA. Such an antenna could be used e.g.\ for the DVB-H system, with proper varactor circuit used to tune the antenna over the frequency range~\cite{Komulainen2007}.


\subsection{Antenna Under Study}

With the meandered PIFA, the fundamental operation of the antenna can be assumed to be similar to normal PIFAs despite the meander line as the patch element. That is, vertical electric field and horizontal magnetic field well below the patch have a well-understandable field distribution while the size is considerably smaller than for a normal PIFA. If the fields are similar to the normal $\lambda/4$ resonating cavity, the fringing electric field is the major field providing radiation.

\begin{figure}[!t]
	\centering
	\includegraphics[width=85mm]{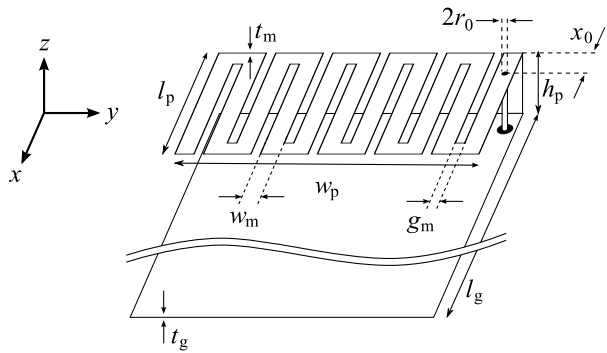}
  \caption{The meandered PIFA dimensions, see also Table~\ref{ta:dimensions}.}
  \label{fig:dimensions}
\end{figure}

In addition to the miniaturized cavity size, we want to study the antenna with a small ground plane, that is typically found in portable devices. The chosen antenna for this study is presented in Fig.~\ref{fig:dimensions} and the dimensions are seen in Table~\ref{ta:dimensions}. The meandered patch and the ground plane form a cavity with the height $h_\lt{p}=8$~mm and volume $V_0=7360$~mm$^3$. The shorting strip is formed when the meander line is turned down and connected to the ground plane. The patch is fed via a feed probe at a distance $x_0$ of a few millimeters to the shorting strip so that the desired matching level is obtained. The ground plane has the dimensions $100 \times 40$~mm$^2$.

The shorting strip and the meander line add up to the complexity and make the use of filling more difficult. The vertical shorting strip has a high current that contributes to radiation. The strong horizontal current at the meander line near the shorting strip may also radiate since the distance from the ground plane is considerable, and also the ground plane is not large enough to produce the image current which would cancel the radiation.

\begin{table}[!t]
    \centering
    \caption{Dimensions for the prototype meandered PIFA of Fig.~\ref{fig:dimensions}.}
    \begin{tabular}{l|l|l}
    Abbr.				&   Value [mm]  &   Explanation \\
    \hline
    $w_\lt{p}$  & $40$	& width of the patch and ground plane \\
    $l_\lt{p}$  & $23$	& length of the patch \\
    $h_\lt{p}$  & $8.0$	& height of the patch from the ground plane \\
    $l_\lt{g}$	& $100$	& length of the ground plane \\
    $t_\lt{g}$  & $0.40$ & thickness of the ground plane \\
    $g_\lt{m}$	&	$1.25$ & gap between the meandered lines \\
    $w_\lt{m}$  & $2.50$ & width of the meander line \\
    $t_\lt{m}$  & $0.20$ & thickness of the meander line \\
    $r_\lt{0}$  & $0.625$ & radius of the feed probe \\
    $x_\lt{0}$  & $\sim 3 \ldots 5$ & distance of the feed from the short circuit \\
    \end{tabular}
    \label{ta:dimensions}
\end{table}

The fields of the meandered PIFA were analyzed in detail using Ansoft HFSS full-wave simulator \cite{HFSS11}. It was seen, that the radiation is still produced mainly by the fringing electric field at the open end of the patch because of the realized omni-directional radiation pattern oriented approximately along the long edge of the ground plane.

\begin{figure}[!t]
  \centering
  \subfigure[]
  {
    \includegraphics[width=85mm]{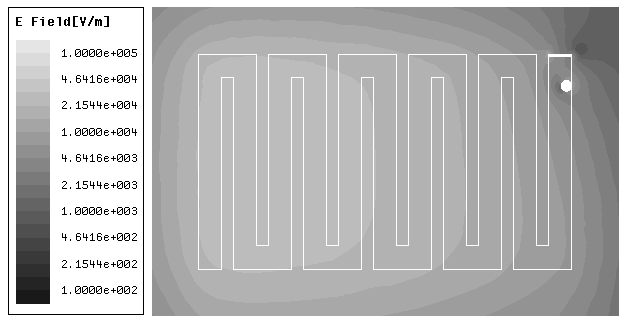}
    \label{fig:ma_mpifa_efield_z3mm}
  }
  \subfigure[]
  {
    \includegraphics[width=85mm]{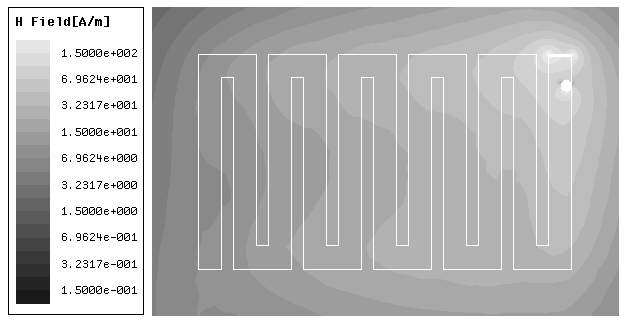}
    \label{fig:ma_mpifa_hfield_z3mm}
  }
  \caption{Simulated a)~electric field strength $|\vt{E}|$ and b)~magnetic field strength $|\vt{H}|$ (90 degress in phase shift) with $1$~W input power for the meandered PIFA at the height $z=3$~mm from the ground plane. The $xy$-plane is seen from the top, and the meandered patch is marked with a solid white line. The shorting strip and the cylindrical feed probe can be seen in the top-right corner of the patch.}
  \label{fig:ma_mpifa_empty_z3mm}
\end{figure}

Fig.~\ref{fig:ma_mpifa_empty_z3mm} shows the resonating electric and magnetic fields inside the cavity in the $xy$-plane, at $z=3$~mm from the ground plane. The vertical $z$-directed electric field is strong at the open end of the patch (the left side of the patch in Fig.~\ref{fig:ma_mpifa_empty_z3mm}), and it is weak near the shorting strip. The magnetic field in the horizontal $xy$-plane, on the other hand, is concentrated on the right-hand side, near the shorting strip. This configuration with predominating horizontal magnetic field is suited for the magneto-dielectric material with in-plane isotropic magnetic response. Also, since the electric field is vertical throughout the cavity, the conductive Fe-N layers do not provide excessive material losses. The fields near the meandered line are strong due to the small details and distances, and the material losses would become too large with the used magneto-dielectric material. We can conclude that the optimal places for the magneto-dielectric filling is near the shorting strip and for the dielectric filling near the open end of the patch (for the latter case, see \cite{Luukkonen2007}).

\subsection{Antenna Measurements}


After the fields in the meandered PIFA were identified and verified by simulations, prototype antennas were manufactured from copper according to the dimensions in Fig.~\ref{fig:dimensions}. Since magnetic loading diminishes the magnetic field, and dielectric does the same to the electric field, the change can also be seen in the ratio of voltage and current ($Z=V/I$), i.e.\ in the ratio of electric and magnetic fields, at the feed position. The dielectric loading decreases $Z$ and the magnetic increases it. Because of this, two antenna prototypes were manufactured, one designed to have a reasonable matching with dielectric and one with magnetic loading.

\begin{figure}[!t]
	\centering
	\includegraphics[width=70mm]{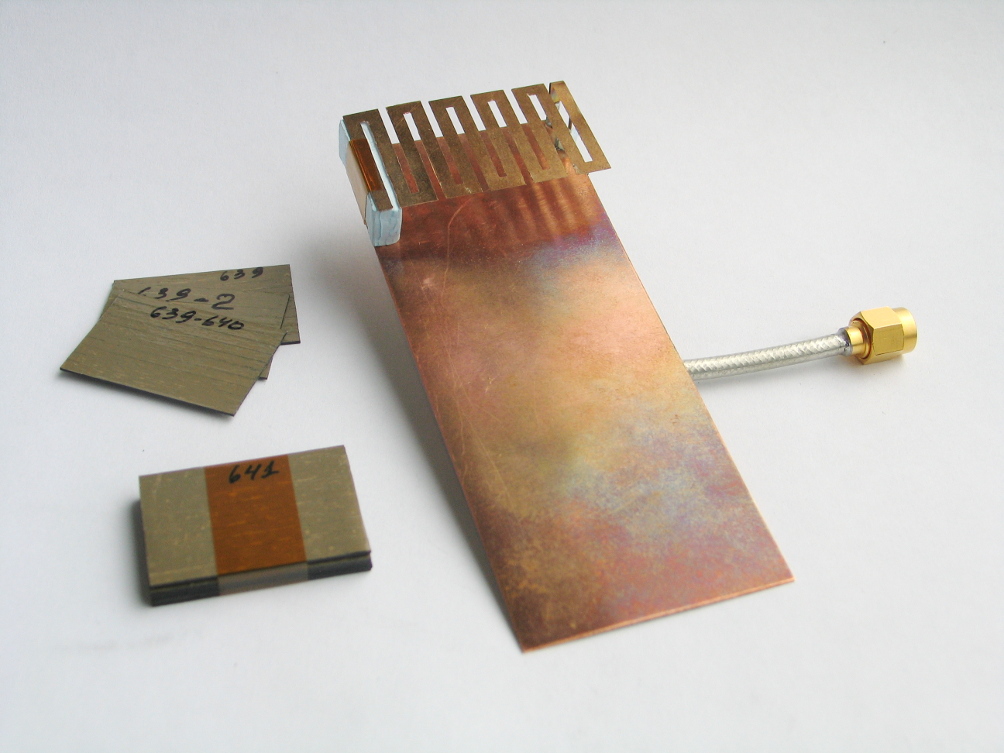}
  \caption{Manufactured antenna with the magneto-dielectric material sheets. The feed cable is brought out from the antenna from the long edge of the ground plane. A piece of Rohacell foam is used to support the open end of the patch.}
  \label{fig:aut_photo}
\end{figure}

The impedance measurements were first done in an anechoic chamber using a vector network analyzer (VNA). A fixed coaxial cable was used to feed the patch as seen in Fig.~\ref{fig:dimensions} and Fig.~\ref{fig:aut_photo} so that the cable protruded from the long edge of the ground plane. This helps to minimize the effect of the measurement cable on the measured impedance. Then, the same impedance measurements were done using the Wheeler cap to find the impedance without radiation~\cite{Wheeler1959}. A cylindrical aluminum cap with 158-mm radius and 148-mm height was used on top of an aluminum ground plane. The antennas were suspended in the center of the Wheeler cap by the fixed cable, and the VNA was connected through the ground plane. The antennas were directed so that the $y$-axis (see Fig.~\ref{fig:dimensions}) pointed towards the bottom of the Wheeler cap. From the measured impedance curves we calculate the quality factor $Q$ using (\ref{eq:QB}) and \ref{eq:QZ}), the radiation efficiency $\eta_\lt{r}$ using (\ref{eq:radeff}) and finally the radiation quality factor $Q_\lt{r}$ using (\ref{eq:Qr}).

\begin{figure}[!t]
  \centering
  \subfigure[]
  {
    \includegraphics[width=85mm]{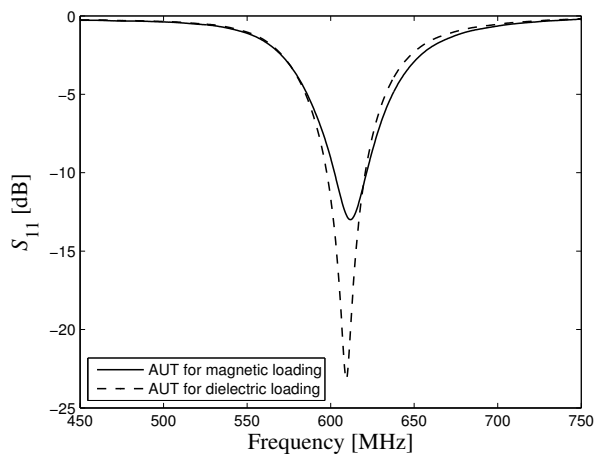}
    \label{fig:aut_empty_s11}
  }
  \subfigure[]
  {
    \includegraphics[width=85mm]{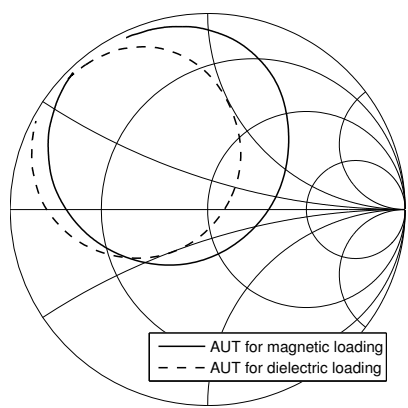}
    \label{fig:aut_empty_smith}
  }
  \caption{The measurement results of $S_{11}$ for the empty manufactured antennas in a)~decibels and b)~in the Smith diagram. The first antenna under test (AUT) is designed for magneto-dielectric loading and the second for dielectric loading.}
  \label{fig:aut_empty}
\end{figure}

The empty antennas were first measured as reference. The realized resonance frequencies were $f_0=611.6$~MHz and $f_0=609.4$~MHz with $\eta_\lt{r}=84.1$\% and $\eta_\lt{r}=85.0$\%, respectively. The difference in $f_0$ is due to slightly different feeding points and manufacturing tolerances of the empty antennas. The simulations run during the design of an empty antenna gave a very similar efficiency of 84.9\% as the measurements and so acted as a verification for the Wheeler-cap measurement setup. The measured reflection coefficients ($S_{11}$) for both of the antennas can be seen in Fig.~\ref{fig:aut_empty}. The reference plane of $S_{11}$ is moved numerically from the connector of the cable to the feeding point of the patch.


\begin{figure}[!t]
  \centering
  \includegraphics[width=85mm]{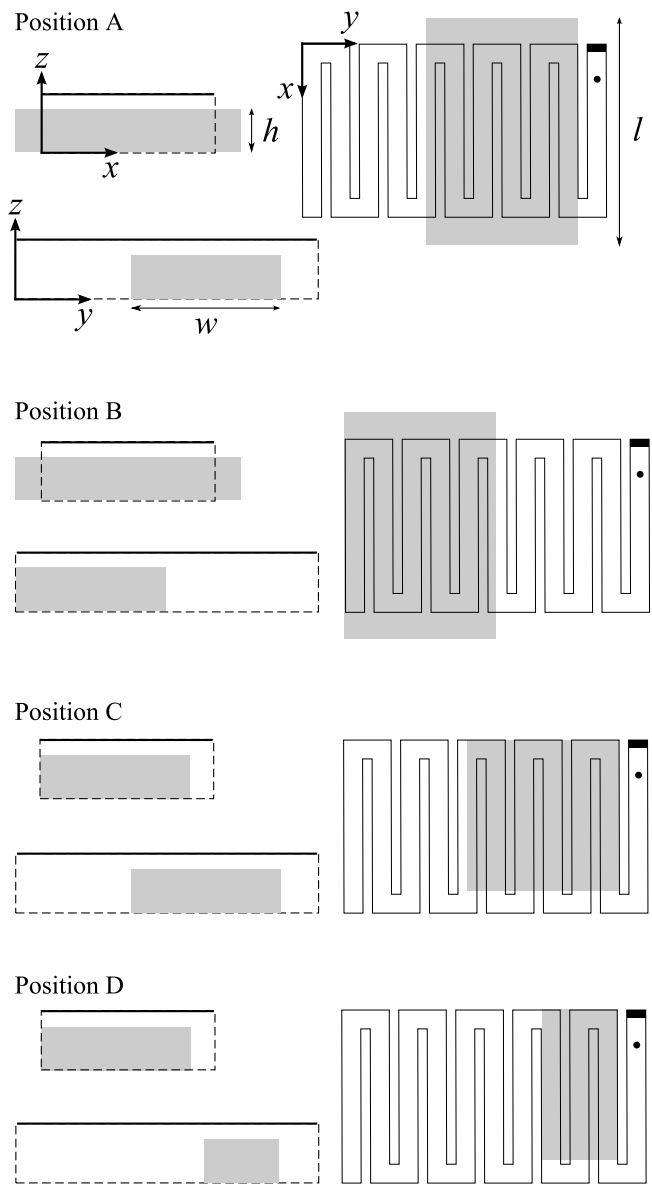}
  \caption{Placements A-D of the magneto-dielectric material. The areas marked with gray show the dimensions of the magneto-dielectric material. The height of the material varies in some cases, see Table~\ref{ta:block_sizes}. Position B is also used as the optimal position for dielectric filling in all reference measurements.}
  \label{fig:placement}
\end{figure}

Next, the measurements for the dielectric and magneto-dielectric filling were conducted as follows. A chosen size and shape of magneto-dielectric material was placed inside the antenna designed for magneto-dielectric loading, and free-space and Wheeler cap impedance measurements were conducted. The positions and sizes are presented in Fig.~\ref{fig:placement} and Table~\ref{ta:block_sizes}. Then, a suitable amount of high-quality Rogers RT/duroid 6002 dielectric substrate was placed in the antenna designed for dielectric loading, in the open end of the cavity where the electric field is highest, since this is the optimal place for dielectric filling~\cite{Luukkonen2007}. The amount and place of the dielectric material was tuned until the same realized resonance frequency $f_0$ was obtained as with the magneto-dielectric material. Now with the same $f_0$ for both antennas, the free-space and Wheeler-cap impedance responses for the dielectric-filled antenna were measured, and $Q$, $\eta_\lt{r}$ and $Q_\lt{r}$ could be calculated for both antennas. The ratio of the radiation quality factors $Q_\lt{r}^\lt{rel} = Q_\lt{r}^\lt{diel}/Q_\lt{r}^\lt{magn}$ was calculated finally for a pair of antenna measurements, where $Q_\lt{r}^\lt{magn}$ and $Q_\lt{r}^\lt{diel}$ are the radiation quality factors with magneto-dielectric and dielectric filling, respectively. $Q_\lt{r}^\lt{rel}$ can be then used to determine which filling material provides the lower quality factor and wider bandwidth potential.

\begin{table}[!t]
    \centering
    \caption{Summary of the magneto-dielectric block sizes used in the measurements.}
    \begin{tabular}{l|l|l|l}
    Position		& $w \times l$ [mm$^2$] &	$h$ [mm]	& $V$ [mm$^3$] \\
    \hline
    Position A	& $20\times30$					& $2$				& $1200$ \\
    Position A	& $20\times30$					& $4$				& $2400$ \\
    Position A	& $20\times30$					& $6$				& $3600$ \\
    Position B	& $20\times30$					& $6$				& $3600$ \\
    Position C	& $20\times20$					& $6$				& $2400$ \\
    Position D  & $10\times20$					& $6$				& $1200$ \\
    \end{tabular}
    \label{ta:block_sizes}
\end{table}

\begin{figure}[!t]
  \centering
  \includegraphics[width=85mm]{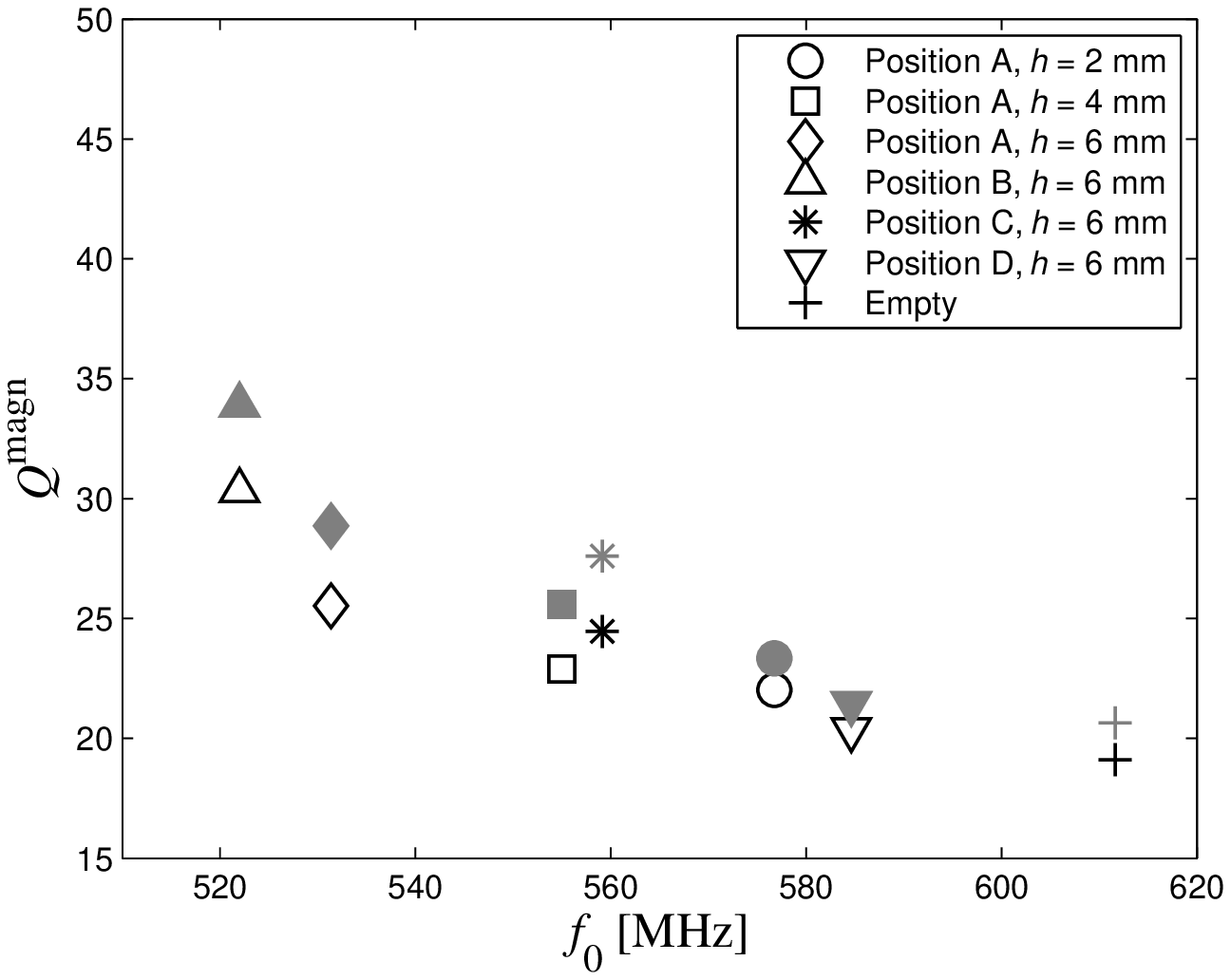}
  \caption{The measured quality factors $Q^\lt{magn}$ for different positions and amounts of magneto-dielectric filling with realized resonance frequencies $f_0$. See Fig.~\ref{fig:placement} and Table~\ref{ta:block_sizes} for the used placements and sizes. The black-and-white markers are calculated using (\ref{eq:QB}) and the gray markers using (\ref{eq:QZ}).}
  \label{fig:Qmagn}
\end{figure}

\begin{figure}[!t]
  \centering
  \includegraphics[width=85mm]{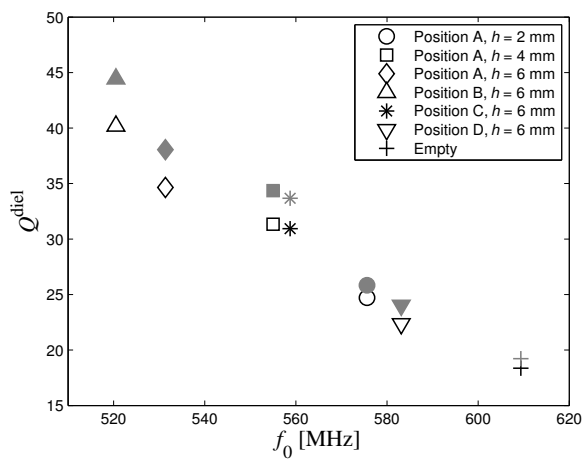}
  \caption{The measured quality factors $Q^\lt{diel}$ for different amounts of dielectric reference filling with the realized resonance frequencies $f_0$ according to $Q^\lt{magn}$ measurements from Fig.~\ref{fig:Qmagn}. The black-and-white markers are calculated using (\ref{eq:QB}) and the gray markers using (\ref{eq:QZ}).}
  \label{fig:Qdiel}
\end{figure}

\begin{figure}[!t]
  \centering
  \includegraphics[width=85mm]{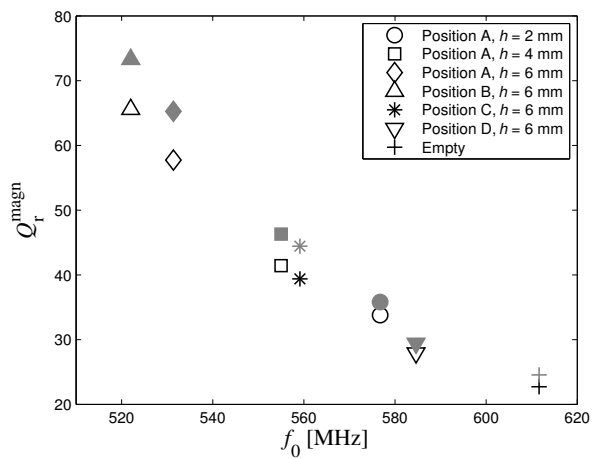}
  \caption{The calculated radiation quality factors $Q_\lt{r}^\lt{magn}$ from Fig.~\ref{fig:Qmagn}. The black-and-white markers are calculated using (\ref{eq:QB}) and the gray markers using (\ref{eq:QZ}).}
  \label{fig:Qrmagn}
\end{figure}

\begin{figure}[!t]
  \centering
  \includegraphics[width=85mm]{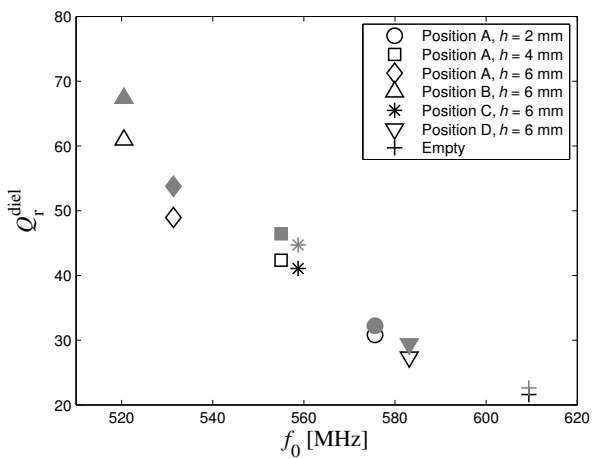}
  \caption{The calculated radiation quality factors $Q_\lt{r}^\lt{diel}$ for different amounts of dielectric reference filling. The black-and-white markers are calculated using (\ref{eq:QB}) and the gray markers using (\ref{eq:QZ}).}
  \label{fig:Qrdiel}
\end{figure}

\begin{figure}[!t]
  \centering
  \includegraphics[width=85mm]{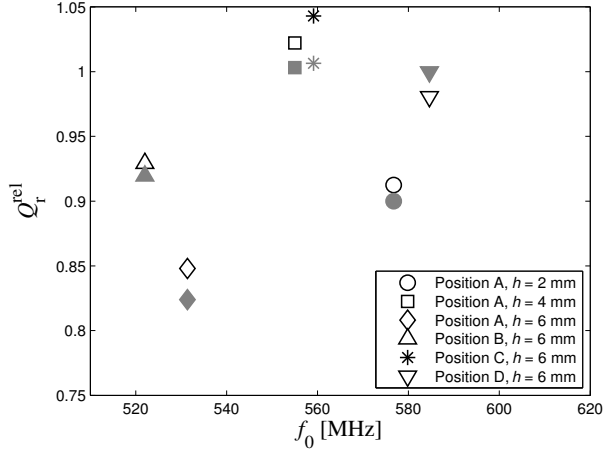}
  \caption{The calculated relative radiation quality factors $Q_\lt{r}^\lt{rel} = Q_\lt{r}^\lt{diel}/Q_\lt{r}^\lt{magn}$. The black-and-white markers are calculated using (\ref{eq:QB}) and the gray markers using (\ref{eq:QZ}).}
  \label{fig:Qrrel}
\end{figure}

The measured quality factors for magneto-dielectric and dielectric fillings, $Q^\lt{magn}$ and $Q^\lt{diel}$, respectively, are presented in Fig.~\ref{fig:Qmagn} and Fig.~\ref{fig:Qdiel}. The values are calculated from $S_{11}$ and $Z_\lt{a}$ using both (\ref{eq:QB}) and (\ref{eq:QZ}), and marked with black-and-white and gray markers, respectively. The requirement of $S_{11}=-6$~dB was used in calculations with (\ref{eq:QB}). The radiation quality factors can be seen in Fig.~\ref{fig:Qrmagn} and Fig.~\ref{fig:Qrdiel}, and their ratio $Q_\lt{r}^\lt{rel}$ in Fig.~\ref{fig:Qrrel}. The measurement results for the radiation efficiencies $\eta_\lt{r}$ and quality factors using (\ref{eq:QZ}) are also presented in Table~\ref{ta:results}.


\begin{table*}[!t]
    \centering
    \caption{Measurement results for the magneto-dielectric (marked as 'magn') and reference dielectric ('diel') filling scenarios. The quality-factor calculations were done using (\ref{eq:QB}).}
    \begin{tabular}{l|l|l|l|l|l|l|l|l|l|l}
    Position and height		& $f_0$ [MHz]	& $S_{11}^\lt{magn}$ [dB]	& $S_{11}^\lt{diel}$ [dB]	& $Q^\lt{magn}$	& $Q^\lt{diel}$	& $\eta_\lt{r}^\lt{magn}$	& $\eta_\lt{r}^\lt{diel}$	& $Q_\lt{r}^\lt{magn}$	& $Q_\lt{r}^\lt{diel}$	& $Q_\lt{r}^\lt{rel}$ \\
    \hline
    Position A, $h=2$~mm	& 576	& -11.0	& -15.6	& 22	& 25	& 0.65	& 0.80	& 34	& 31	& 0.91  \\
    Position A, $h=4$~mm	& 555	& -10.3	& -11.1	& 23	& 31	& 0.55	& 0.74	& 41	& 42	& 1.02  \\
    Position A, $h=6$~mm	& 531	& -9.5	& -10.1	& 26	& 35	& 0.44	& 0.71	& 58	& 49	& 0.85  \\
    Position B, $h=6$~mm	& 521	& -8.8	& -8.6	& 30	& 40	& 0.46	& 0.66	& 66	& 61	& 0.93  \\
    Position C, $h=6$~mm	& 559	& -9.4	& -11.3	& 24	& 31	& 0.62	& 0.75	& 39	& 41	& 1.04  \\
    Position D, $h=6$~mm  & 584	& -11.7	& -16.6	& 20	& 22	& 0.73	& 0.82	& 28	& 27	& 0.98  \\
    \end{tabular}
    \label{ta:results}
\end{table*}

The fixed feed cable, which was used to make reliable impedance and efficiency measurements, has however an effect on the impedance. Measured $Q^\lt{magn}$ and $Q^\lt{diel}$ include a small uncertainty due to additional stored energy in the cable and due to affecting the stored and radiating field distributions of the antennas.
%
Some inherent uncertainty appear also from the VNA's $S$-parameter measurement. The error on the radiation efficiencies is estimated to be around $\pm 2$\% with small miniaturization and relatively low losses, and around $\pm 5$\% with the highest measured miniaturizations. When calculating and comparing the $Q$ values using (\ref{eq:QZ}), a constant bandwidth requirement of $S_{11}=-6$~dB was used. However, the measured reflection coefficients at the center frequency were not constant with different antennas and filling scenarios (see Table~\ref{ta:results}). This has also a small uncertainty effect on $Q$.
%
The two antennas under comparison were not perfectly identical due to manufacturing constraints, e.g.\ there may be some differences in the dimensions. Especially, even the smallest variations in the patch height affects the measured resonance frequency, as do small variations in the feed position of the probe.



\section{Discussion }
\label{sec:discussion}

The measured $Q^\lt{magn}$ and $Q^\lt{diel}$ in Fig.~\ref{fig:Qmagn} and Fig.~\ref{fig:Qdiel}, respectively, behave as expected. With the magneto-dielectric filling the values are considerably lower due to higher dissipative losses than with the pure dielectric filling. As the effect of dissipative losses is removed, we see the effect of the material response to the quality factors. The measured radiation quality factors in Fig.~\ref{fig:Qrmagn} and Fig.~\ref{fig:Qrdiel} are similar. In both cases we see the effect of miniaturization: as the resonance frequency decreases, the quality factor also increases. This underlines the fact that with small antennas the resonance frequency must be the same for fair comparison. The recently proposed equation (\ref{eq:QZ}) seems to give reasonable estimates for $Q$. The results from (\ref{eq:QB}) and (\ref{eq:QZ}) are within uncertainty limits from each other. With the used antenna type the magnetic response does not lead to bandwidth-independent miniaturization, as it would be expected for e.g.\ $\lambda/2$ patch antennas \cite{Ikonen2008,Karilainen2009}, but the differences are rather small.

When we examine the individual filling cases of Table~\ref{fig:placement}, we see the differences for $Q_\lt{r}$ from Fig.~\ref{fig:Qrrel} and Table~\ref{ta:results}.
The material blocks used in position A at different heights are positioned in strong magnetic field. As the height of the block increases, $f_0$ decreases, as does $\eta_\lt{r}$. The 4-mm tall magneto-dielectric block and reference dielectric fillings lead to approximately the same $Q_\lt{r}$ as $Q_\lt{r}^\lt{rel}$ is close to unity, but as $h$ increases, the magnetic losses also increase and the reference dielectric provides better $Q_\lt{r}$.
Magneto-dielectric filling in position B, on the other hand, miniaturizes the antenna mostly due to the electric field effect. The reference material has a pure dielectric response, whereas the magneto-dielectric material has both material-type responses. One would expect that the radiation quality factors would produce similar results, but the measured $Q_\lt{r}^\lt{rel}<1$. The reason behind this is probably in the different radiation properties due to different field distributions inside the cavity.
Positions C and D give similar results to position A, only with a smaller footprint of the filling block. We see that the block in position C behaves similarly as the block in position A with $h=4$~mm and the same volume, and provides $Q_\lt{r}^\lt{rel}$ close to unity. However, the block of position D utilizes the strong magnetic field close to the shorting strip more efficiently than the block with $h=2$~mm in position A, leading to higher $Q_\lt{r}^\lt{rel}$.


Does the magneto-dielectric material perform better in terms of lower radiation quality factor in the studied meandered PIFA? Due to the discussed uncertainties, the observed $Q_\lt{r}^\lt{rel}$ above unity in some cases in Fig.~\ref{fig:Qrrel} are within the uncertainty limits, so no definitive answer can be given.
The observed decrease in the resonance frequency, i.e.\ miniaturization, was moderate (8\%) for the filling cases that benefit most from the magneto-dielectric material. This suggests that with a more effective material placement and higher values of $\mu_\lt{r}$, the effect would be stronger.
With the used magneto-dielectric material, higher values of $\mu_\lt{r}$ (higher filling ratio of Fe) would also lead to higher material losses, which in turn would decrease the radiation efficiency. Therefore, with available magnetic materials, we cannot use only $Q_\lt{r}^\lt{rel}$ as the only figure of merit, but in addition the realized radiation efficiency of the miniaturized antenna must be high enough for practical applications.
However, we must take into account that the material losses most likely have some effect to the measured $Q_\lt{r}$ through changed field distributions inside the antenna. That is, after the power has been accepted to the antenna volume, the radiated power may still depend e.g.\ on the filling material and types.

It can be seen from the measurement results that when filling a small antenna partially, positioning of the filling material is crucial for achieving the desired effect. The used magneto-dielectric material dictates additional restrictions for the antenna. Also the losses depend on the field configuration. Even small material losses, in addition to the conductor losses, can lead to greatly diminished radiation efficiency of small antennas.

If it would be possible to manufacture materials with adequate magnetic response and low losses in the gigahertz-range, the situation would be different. Since the radiation quality factor of small antennas with fixed volume for hand-held devices decreases as frequency increases, such antennas would tolerate larger losses at higher frequencies.


\section{Conclusion}
\label{sec:conclusion}

When measuring and comparing small antennas with material loading, special care must be paid to make the comparison fair. The radiation quality factor as the figure of merit for small resonant antennas and the most beneficial partial filling places for magneto-dielectric and dielectric material samples in resonant antennas were discussed. Measurement results were presented for a meandered PIFA with fabricated magneto-dielectric material and compared according to the presented guidelines to a reference dielectric material filling. It was seen that the benefit from the magneto-dielectric material is strongly dependent on the positioning of the material. The measured benefit is moderate due to low miniaturization ratio with the used antenna type and small volume of the partial filling and also due to magnetic losses. Magnetic materials with higher values of weakly-dispersive permeability and with smaller losses are needed for effective miniaturization of antennas.


\section*{Acknowledgments}

This research was funded in part by Intel Corporation and Nokia Corporations. A.~O.~Karilainen, C.~R.~Simovski and S.~A.~Tretyakov acknowledge the support of the Academy of Finland and Nokia through the Center-of-Excellence program. A.~Karilainen would like to thank Mr.~E.~Karha and Mr.~L.~Laakso for technical assistance. A.~N.~Lagarkov, S.~A.~Maklakov, K.~N.~Rozanov, and S.~N.~Starostenko are grateful to the RFBR for partial support of the study according to the grant no. 09-08-01161.


\bibliographystyle{../../../bibtex/IEEEtranBST/IEEEtran}
\bibliography{../../../bibtex/IEEEtranBST/IEEEabrv,../../../bibtex/karilainen}

\end{document}